\documentclass[sigconf, nonacm]{acmart}
\usepackage[english]{babel}
\usepackage{amsmath}
\usepackage{graphics}

\AtBeginDocument{%
	\providecommand\BibTeX{{%
			\normalfont B\kern-0.5em{\scshape i\kern-0.25em b}\kern-0.8em\TeX}}}

\begin{document}
	\title{Synchronous Consensus in Partial Synchrony}
	
	\author{Ivan Klianev}
	\email{Ivan.Klianev@gmail.com}
	\affiliation{%
	\institution{Transactum Pty Ltd}
	\city{Syndey}
	\country{Australia}
	}
	\begin{abstract}
	We demonstrate a deterministic Byzantine consensus algorithm with synchronous operation in partial synchrony. It is naturally leaderless, tolerates any number of $ f<n/2 $ Byzantine processes with 2 rounds of exchange of originator-only signed messages, and terminates within a bounded interval of time. The algorithm is resilient to transient faults and asynchrony in a fraction of links with known size per number of faulty processes.  It circumvents asynchronous and faulty links with 3-hop epidemic dissemination. Key finding: the resilience to asynchrony of links and the enabled by it leaderless consensus in partial synchrony ensure algorithm operation with simultaneous validity, safety, and bounded liveness.	
	\end{abstract}

		

	\maketitle	
	
	\section{Introduction}
	 	
	 	The theoretical divergence between consensus in synchrony \cite{PeaseShostakLamport80} and in partial synchrony \cite{DworkLynchStockm1988} is to a certain degree not natural. Point of separation is a single assumption, based on the definition of communication synchrony \cite{DworkLynchStockm1988}, which allows tuning per needs.
	 	Synchronous consensus can tolerate crash fails \cite{PeaseShostakLamport80}, Byzantine faults \cite{PeaseShostakLamport80} \cite{DolevStrong1983}, and faulty links \cite{Dolev1981}, but not asynchronous links. This is despite the observations that asynchronous links have the same effect on synchronous consensus as the faulty links.
		
		We solved the impossibility of bounded consensus termination in partial synchrony. The number of tolerated asynchronous links is a function of the number of faulty processes per system size.	The asynchronous links are circumvented with 3-hop epidemic dissemination \cite{BirmanEtAl1999} and thus the time for delivery is 3 times longer, yet bounded and known in advance. Critical for the solution is its consensus algorithm where the number of communication rounds, needed for resilience of synchronous deterministic protocol with authenticated messages \cite{RivestShamirAdleman78} to \textit{f} Byzantine processes, is smaller than the known minimum of $ (f+1) $ communication rounds \cite{DolevStrong1983}. The algorithm uses only 2 communication rounds regardless of the number of tolerated Byzantine processes. With more than 12 processes in partial synchrony, it terminates faster than a synchronous Byzantine \cite{DolevStrong1983} system with the same size and delivery bound. Moreover, it makes leaderless consensus \cite{BoranSchiper_2010} \cite{AntoniadisEtAl2021} in partial synchrony a no more existing problem.

	\subsection{Background}	
		
		The term \textit{partial synchrony} reflects the practice-driven
		attention to messaging that is neither fully asynchronous nor fully synchronous. As a theoretical field, it can be viewed as two separate spaces with a borderline blurred by the conditionality of tolerated asynchrony.
		

		\textit{Unmanageable} space: Until recently the only approach in dealing with asynchrony was the model of \textit{eventual synchrony} \cite{DworkLynchStockm1988}. It reflects the
		reality where intervals of non-infringed synchrony with unknown duration
		are followed by intervals with infringed synchrony. It accepts the inevitability of infringed synchrony and allows termination only during large enough intervals with non-infringed synchrony \cite{DuttaGuerraoui_2002} \cite{Alistarh_2008}.
		Yet, a prepublication \cite{Klianev_FLP} shows solving consensus in complete asynchrony under the model used for proving the FLP impossibility result \cite{FLP_result}, on condition of non-dependence of agreement on the content of the initial values.

		\textit{Manageable space}: The industry-preferred model \cite{CastroLiskov2002PBFT} is to rely on a consensus leader rather than wait for eventual synchrony. With this model, the dependence on asynchrony is replaced with dependence on a possibly Byzantine \cite{LamportShostakPease82} single point of failure. 
		The use of cryptographically authenticated messages \cite{RivestShamirAdleman78} prevents Byzantine processes from doing more harm on consensus safety than the crash-failed processes and allows solving consensus with a simple majority of processes \cite{PeaseShostakLamport80} \cite{DolevStrong1983}. Nevertheless it cannot prevent the ability of a  Byzantine leader to spoil liveness.

		None of the above models ensures bounded termination in the presence of asynchrony. The termination either depends on external factors beyond control or on a possibly Byzantine leader that can delay the broadcasting of its proposing message without being detected. We apply gossiping \cite{BirmanEtAl1999} \cite{EugsterEtAl2004}, normally used to circumvent faulty and non-existing links, for bounded delivery in presence of asynchrony and show that leaderless consensus and bounded termination are natural outcomes from the ability to mitigate some amount of asynchrony.	
	
		\subsection{Motivation}

		The decentralised forms of distributed data systems operate with global replication of local transactions and atomic
		\cite{Herlihy_Wing_1990}
		consistency across local systems via state machine replication
		\cite{Lamport_StateMachine} 
		\cite{Schneider_1990} 
		\cite{Lampson_1996}.
		Agreement for the purposes of state machine replication is about an ordered set of initial data and its actual values cannot make consensus impossible \cite{Klianev_FLP}. Systems with no single point of failure and bounded liveness could solve the vulnerability \cite{BernsteinHadzilacosGoodman} of distributed transactions \cite{BruceLindsay79} \cite{JimGray78} \cite{LampsonSturgis76} via being coordinated directly by its participants with a time-controlled workflow \cite{Garcia-Molina_Salem_1987}. 

		Moreover, the possibility for bounded termination of consensus in partial synchrony is expected to trigger attention on the so far axiomatic acceptance that a consensus protocol cannot operate with ensured simultaneous validity, safety, and liveness in partial synchrony \cite{GilbertLynch_2012}, and even less with a bounded liveness.
		
	\subsection{Problem Formulation}
		
		We have to show solving a problem with the following formulation:
		
		\textbf{Given}: i) consensus system $ S $ of $ n $ processes, $ n > 2f $, where $ f $ is the tolerated number of crash-failed and Byzantine processes; ii) communicate using authenticated messages; iii) every process $P_i$, where $i\in\{1,...,n\}$, has an initial value $v_i$; iv) set $ C \subseteq S $ comprises no less than $ (n-f) $ correct processes; v) every $P_{i} \in C$ delivers own message $m_i$ to every other $P \in C$ over synchronous links with up to 3 hops during periods with asynchronous and/or faulty links; and vi) terminates with set of processes $ C'$, $ (C' \subseteq C, |C'| \ge (n-f)) $.
		
		\textbf{Demonstrate}: ability to ensure that all processes in $ C' $ agree on an $n$-vector $ V=(v_1,...,v_n)$ that contains the initial values of at least all processes in $ C' $, and less than $ f $ vector elements are replaced with $ \varnothing $ denoting an empty element, with 2 rounds of exchange of messages and within a time interval $ T_T \leq 2*3* T_S$, where $T_S$ is the time bound for synchrony of links.

	\subsection{Contribution}

		Synchronous deterministic Byzantine consensus algorithm with:

		1. Quantified tolerance to links asynchrony and faults.
		
		2. $ f $-resilience, $ f < n/2 $, implemented with 2 messaging rounds.
		
		3. Ensured simultaneous validity, safety, and bounded liveness.
		
		4. Naturally leaderless operation.

	\subsection{Similar Works}
	
		\textbf{Synchronous Consensus with Transient Faults}
		
		Santoro \textit{et al} \cite{SantoroEtAl1989} studies the possibility for agreement in a fully synchronous system and presents conditions making the agreement impossible. It shows that $(n/2 + 1)$ binary agreement is impossible if, at each time unit, there is one process whose all $(n-1)$ messages are lost. 
		This result applies to agreements that depend on the content of the initial values. It does not apply \cite{Klianev_FLP} to our agreement, which is about ordered initial values.
	
		\textbf{Synchronous Consensus Over a Fraction of Links}
		
		Siu \textit{et al} \cite{Hin-SingSiuEtAl1998} presents a concept for fault-tolerant virtual channels in a synchronous deterministic Byzantine consensus system, where processes and links can fail independently and simultaneously. It ensures reliable delivery for the purposes of synchronous consensus over a fraction of links but does not consider links asynchrony.
		
		Aguilera \textit{et al} \cite{AguileraEtAl2003} \cite{AguileraEtAl2004} demonstrate a case of solved synchronous deterministic Byzantine consensus with a small fraction of all links operating synchronously. The shown result – one correct process having all inbound and outbound links synchronous – 
		is with a negligibly small probability to happen and it is not a solution.

		\textbf{Consensus via Gossip-based Reliable Multicast}	
		
		In a distributed system where every process knows every other, the algorithm of Birman \textit{et al} \cite{BirmanEtAl1999} uses gossips, a.k.a. epidemic dissemination of messages, to deliver the messages to all recipients, even if the sender fails before sending to all. The algorithm tolerates process crash failures and a number of failures of links. Yet it not about Byzantine failures, asynchrony of links, or bounded delivery. 
		
		\textbf{Atomic Broadcast for Synchronous Replicated Storage}
		
		Cristian \textit{et al} \cite{FlaviuCristian_EtAl_1995} demonstrates ability to handle random communication delays and link failures, using multi-hop delivery in presence of crashed-failed and Byzantine processes, with objective of  \textit{information diffusion} for the purposes of synchronous replicated storage. 
		Even so, it does not show ensured bounded termination. In contrast, we use multi-hop delivery to mitigate the effect of asynchronous and/or faulty links and ensure bounded termination.
		
	\subsection{Content}		
	
		Section 2 presents the system model. Section 3 presents the solution. Section 4 presents the formal proofs. Section 5 summarizes the findings. Section 6 concludes the paper. The Appendix quantifies the interdependence between the actual number of faulty processes and the number of tolerated asynchronous and faulty links.
	
	
	\section{System Model}
		
		
	\subsection{Assumptions}	
		
		About:

		\textbf{Processes}: Synchronous processes, executing on CPUs with a fixed upper bound on processing cycle speed difference, having access to fully synchronised clocks with no drift, and starting consensus rounds in a fully synchronised manner with a negligibly small lag.

		\textbf{Communication}: The processes communicate by sending each other messages, over a majority of links with synchronous performance and possible at any point of time existence of a random minority of links performing asynchronously.

		\textbf{Messages}: The messages are signed by their creators with third party verifiable signatures, i.e. every process can verify the authenticity of a signed message as every process knows the public key of every other process' private key for signature.

		\textbf{Tolerance}:	Within the limits, tolerance to simultaneous: i) link failures; ii) asynchrony of links; iii) crash-failed processes; and iv) Byzantine processes, i.e., a process with arbitrary behaviour expressed as omission to send to all recipients, delay before sending to one or more recipients, and deceit in regard to message content.

		\textbf{Tolerance Limits}:	A plain majority of correct processes with bounded indirect delivery between themselves exists at any time with no reliance on asynchronous links becoming synchronous, or faulty links becoming correct, within a consensus round.
	
	\subsection{The Model}
		
		A consensus protocol operates with a system of $ n $ process $(n \geq 3, n >2f)$. where $ f $ is the number of tolerated crash-failed and/or Byzantine-faulty processes. Each process $P$ has an \textit{input register} $R^{In}$, an \textit{output register} $R^{Out}$, and an unbounded amount of internal storage. Register $R^{Out}$ is a vector of $n$ elements $R^{In}$:

		$R^{Out}=(	R^{In}_{1},R^{In}_{2},...,R^{In}_{n} )$
		
		\textit{Internal state} of a process $ P $ comprises value in $R^{In}$ and value in $R^{Out}$, together with a program counter and internal storage.

		\textit{Initial values} of the internal state are fixed, with an exception for the initial value in $R^{In}$. More specifically, $ R^{In} $ starts with a \textit{length prefixed string} having an arbitrary value from a finite universe, and $ R^{Out} $ starts always with the fixed value $R^{Out}=(	\varnothing,\varnothing,...,\varnothing )$,	where $ \varnothing $ is a null marker denoting an empty vector element.
		
		Objective of the consensus protocol is to establish agreement about the value of $R^{Out}=(	R^{In}_{1},R^{In}_{2},...,R^{In}_{n} )$ between $\ge (n-f)$ correct processes with exchange of two type of messages:
		
		- \textit{Initial Value} message of $ P_i $ contains the initial value in $R^{In}_i$;

		- \textit{Agreement Proposal} message of process $ P_i $ is  vector $ A^{Prop}_i $ of $ n $ elements $ R^{In}_j $, where $ j \in\{1,...,n\} $ and $\le f $ elements have value $\varnothing $.

		A \textit{message} is a tuple $(P_{s}, P_{d}, m)$, where $P_{s}$ is the 'source' process,  $ P_{d} $ is the 'destination process', and $ m $ is a 'message value' from a fixed universe $ M $. The message system maintains a \textit{message buffer} for each one-way link $ L_{pq} $ that delivers messages from  process $P_{p}$ to process $ P_{q} $.
		Message buffer is a multi-set of messages that have been sent but not yet delivered, in case it models the behaviour of a correct synchronous or a correct asynchronous link. It is an empty set of messages in case it models a faulty link. The message system supports two abstract operations:
		
		- \textit{Send}($P_{s}, P_{d}, m$). It places ($P_{s}, P_{d}, m$) in a message buffer.
		
		- \textit{Receive}($P_{d}$). It deletes some message ($P_{s}, P_{d}, m$) from the buffer and returns $ m $, in which case ($P_{s}, P_{d}, m$) is \textit{delivered}, or $ \varnothing $ and leaves the buffer unchanged.
		
		A message buffer representing a link $ L_{pq} $ operates with a delay made by two components: one representing the time necessary for a message to physically travel from process $P_{p}$ to process $ P_{q} $; and another representing the delay resulting from congestions on the physical link. The delay operates per message by loading a counter with the delay value on every message entering the buffer. \textit{Receive}($P_{q}$) does not return $ \varnothing $ only if among all buffers representing links to process $ P_{q} $ there is a message with expired delay. Thus, the message system acts deterministically. Every message ($P_{p}, P_{q}, m$) in the message buffer is delivered, subject only to the condition that \textit{Receive}($P_{d}$) is performed until expiration of delay $ D_{pq} $ associated with link $ L_{pq} $ at the moment the message had entered the buffer. Link $ L_{pq} $ is considered synchronous if $ D_{pq} $ is not larger than the delivery time considered synchronous and asynchronous otherwise.
		
		\textit{Message Event} is a tuple ($P_{s}, P_{d}, m$) indicating that $ m $, having content created by process $P_{s}$, has been delivered to process $P_{d}$.
		
		\textit{Time Event} is a trigger for execution of a sequence of activities related to preparation and broadcasting of a protocol message.
	
		\textit{Decision value} is a vector $V^{De}$ of $ n $ elements that contains no more than $ f $ elements with value $\varnothing $. \textit{Decision state} is an internal state a process $ P $ reaches after $ P $ knows that no less than $ (n-f) $ processes, including itself, will propose exactly the same $A^{Prop}$ for decision value $V^{De}$. Once $ P $ has reached a decision state, $A^{Prop}$ becomes $V^{De}$ and $P$ writes $V^{De}$ in its output register $ R^{Out} $.

		$ P $ acts deterministically according to a transition function $ \Phi $, which cannot change the value of $R^{Out}$ once $ P $ reaches a decision state. The system is specified by $ \Phi $ associated with each of the processes and the initial values in the input registers. The correct processes strictly follow $ \Phi $. The Byzantine processes do not.

		\textit{Configuration} is a set that contains the internal state of every process of the system and the contents of its message buffer.

		\textit{Step} is what takes one configuration to another and consists of a primitive step by a single process $ P $. A step executes a sequence of activities as one atomic operation. Execution of step can be triggered by a \textit{message event} or by a \textit{time event}. Let $ C $ be a configuration. A message-triggered step occurs in two phases. First, \textit{Receive}($P$) on the message buffer in $ C $ obtains a value $m \in M \cup |\varnothing|$. Next, depending on the internal state of $ P $ in $ C $ and on value $ m $, process $ P $ may enter a new internal state. Since the processes are deterministic, a message-triggered step of $ P $ is determined by the internal state of $ P $ and the received value $ m $. A time-triggered step prepares and sends a finite set of messages to other processes. It is determined by the internal state of $ P $ at the point of time when the \textit{time event} has happened.

		Other terms:

		- \textit{Schedule}. A schedule from configuration $ C $ is a finite sequence of events that can be applied in turns, starting from $ C $. For a process $ P $ schedules are commutative \cite{FLP_result}.

		- \textit{Run}. A sequence of steps taken in association with a schedule.

		- \textit{Correct process}. A process is correct in a run if it takes infinitely many steps and strictly follows function $ \Phi $.
		
		- \textit{Crash-failed process}. A process that executes correctly until unexpectedly stops and once stopped, cannot restart.
		
		- \textit{Byzantine-faulty process}. A process that deliberately deviates from the strict following of function $ \Phi $.

		Configurations:

		- \textit{Initial}. An initial configuration is one in which each process starts at the initial state and the message buffer is empty.

		- \textit{Resulting}. A configuration that is a result from execution of any schedule, regardless of the number of events.

		- \textit{Reachable}. A configuration is reachable if it is a result from execution of a schedule with a finite number of events.

		- \textit{Accessible}. A configuration is accessible if it is reachable from some initial configuration.

		- \textit{Decision}. A decision configuration is an accessible configuration where at least $ (n-f) $ processes have reached a valid decision state.
		
		- \textit{Valid decision state}. A vector with size $ n $ that contains the \textit{initial values} of no less than the agreed processes as vector elements and $ \varnothing $ marker as value in the rest of vector elements.
		
		Run types:

		- \textit{Admissible run}. A run starting from an initial configuration where at most $ f $ processes are not correct and all messages sent within a set of no less than $ (n-f) $ correct processes are delivered with up to three hops over synchronous links.

		- \textit{Deciding run}. An admissible run that has reached a decision configuration. 

		Correctness Condition:

		- \textit{Correct protocol}. A consensus protocol is correct only if every admissible run is a deciding run with one decision state.

	\section{The Solution}

		The consensus algorithm consists of two parts: i) indirect delivery; and ii) synchronous Byzantine consensus protocol with agreement about ordered initial values \cite{PeaseShostakLamport80}.

	\subsection{Indirect Delivery}

		A cryptographically signed \cite{RivestShamirAdleman78} message might be relayed multiple times before reaching its final destination, in effect as if received directly from its creator. So, a message might reach its recipients via \textit{indirect channels} – virtual aggregations of multi-hop indirect delivery paths.

		\textbf{Indirect Delivery Paths}

		Consensus system $S$ of $n$ processes $(n \geq 2)$ has $n*(n-1)$ direct channels $c_{pq}$ $(1 \leq p \leq n, 1 \leq q \leq n, p \neq q)$. Channel $c_{pq}$ is a one-way link for delivery from process $P_{p}$ to process $P_{q}$. 

		With $(n \geq 3)$, $S$ has also a set of two-hop indirect
		delivery paths comprising $n*(n-1)*(n-2)$ individual paths, as every direct path has $(n-2)$ two-hop alternatives. 
		A message from process $P_{p}$ to process $P_{q}$ can be delivered with two hops: the first hop from $P_{p}$ to process $P_{r}$ $(1 \leq p \leq n, 1 \leq q \leq n, 1 \leq r \leq n, p \neq q,  r \neq p,  r \neq q )$, and the second hop from process $P_{r}$ to $P_{q}$. Indirect delivery over two links with bounded delivery is longer but still bounded.

		With $(n \geq 4)$, system $ S $ has as well a set of three-hop indirect delivery paths comprising $n*(n-1)*(n-2)*(n-3)$ paths as every \textit{two-hop indirect} path has $(n-3)$ three-hop alternatives.
		A message from process $P_{p}$ to process $P_{q}$ can be delivered with three hops: the first from $P_{p}$ to process $P_{r}$, the second from $P_{r}$ to process $P_{s}$ $(1 \leq p \leq n, 1 \leq q \leq n, 1 \leq r \leq n, 1 \leq s \leq n, p \neq q,  r \neq p,  r \neq q, s \neq p, s \neq q, s \neq r )$, and the third from process $P_{s}$ to process $P_{q}$. 

		\textbf{Indirect Delivery Channels}

		An indirect delivery path depends on strict and timely execution of a relaying algorithm by the involved processes. 
		Indirect channel is a virtual construction of links and processes that interact according to a relaying protocol for circumvention of a faulty or asynchronous link with a longer but still bounded delivery. A set of \textit{2-hop indirect channels} operates with one process that helps deliver with two hops. The helper process multicasts a message received from sender to $(n-2)$ recipients, thus enacting $(n-2)$ channels; A set of \textit{3-hop indirect channels} operates with $1 +(n-2)$ helper processes. The first helper enacts $(n-2)$ \textit{2-hop indirect channels}. The group of $(n-2)$ second helpers enacts the third hop where each of the second helpers multicasts the received message to $(n-3)$ recipients, on its first receiving. Indirect channels are formed dynamically.

	\subsection{Synchronous Consensus Protocol}	

		It operates in two phases and exchanges two types of messages:

		\textbf{Phase One}

		At phase start, every process broadcasts an \textit{Initial Value} message. During the phase, every process receives \textit{Initial Value} messages from the rest of the system processes and assembles a \textit{dataset} vector. At phase end, every process computes an \textit{agreement proposal} vector, using the data accumulated and ordered in the \textit{dataset} vector.

		\textbf{Phase Two}

		At phase start, every process broadcasts an \textit{Agreement Proposal} message. During the phase, each process receives the same type of messages from the rest of the system processes and assembles a vector where each element is a received \textit{agreement proposal} vector. At phase end, every process computes a \textit{received proposals} matrix using data from the assembled vector of vectors.

		\textbf{The Intuition}

		At the end of Phase Two every process, in the largest majority of correct processes with bounded delivery between themselves, knows the \textit{initial value} and the \textit{agreement proposal} of every other process in that majority. This information is sufficient for every process in the majority to decide exactly the same valid agreement value at exactly the same point of time.

		\textbf{Constant Number of Messaging Rounds}

		Unlike the authenticated algorithm for Byzantine agreement of Dolev and Strong \cite{DolevStrong1983}, we use signed \textit{Agreement Proposal} messages, so that the \textit{Initial Value} messages do not have to carry $ (f+1) $ signatures. Thus we made the Exponential Information Gathering \cite{Amotz_Bar-Noy_EtAl_1987} irrelevant to our algorithm. Its resilience to any number of tolerated Byzantine processes requires two communication rounds. 

	\subsection{Protocol Messages}	

		The algorithm operates using two message types, authenticated with asymmetric signature of the original sender:

		\textbf{\textit{Initial Value} Message}

		\textit{Initial Value} message of process $ P_i $ contains a tuple

		$ IV_i = (R^{In}_i, S^{In}_i )$, where $ S^{In}_i $ is signature of $ P_i $ over the content of its initial value $ R^{In}_i $.

		\textbf{\textit{Agreement Proposal} Message}

		\textit{Agreement Proposal} message of process $ P_i $ is a tuple 

		$ AP_i = (A^{Prop}_i, S^{Prop}_i) $, where 

		$A^{Prop}_i = (hIV_{i1}, hIV_{i2}. ..., hIV_{in}) $ is a vector of $ n $ elements, each element being a tuple 

		$ hIV_{ij} = (hR^{In}_{ij}, S^{In}_{ij} ) $ and $\le f $ elements replaced with value $\varnothing $, 

		where $ hR^{In}_{ij} $ is a hash digest over the value $ R^{In}_{ij} $, which is the \textit{initial value} $ R^{In}_j $ of process $ P_j $ received by process $ P_i $,

		where $ S^{In}_{ij} $ is the signature of $ P_j $ over the content of \textit{initial value} $ R^{In}_j $ received by process $ P_i $, and

		where $ S^{Prop}_i $ is the signature of process $ P_i $ over the content of $ A^{Prop}_i $.

	\subsection{The \textit{Trim} Rule} 

		Let $ S $ be a system of $ n > 2f $ processes, which tolerates $ f $ faulty processes, and set $ C $ be the largest in S set of correct processes with bounded delivery within the set. Every process in a set comprising $ \ge (n-f $) processes, with bounded indirect delivery within the set, has a reasonable ground to consider that it is in $ C $. Yet multiple sets might satisfy this requirement without being \textit{the} set $ C $.

		The protocol requires every correct process that is in a majority set to decide about the agreement value. If process $ P $ is in the largest majority set $ C $, then $ \ge (n-f) $ processes must decide the same agreement value. Yet $ P $ being in a majority set $ D $, $ |D| \ge (n-f) $, does not know whether $ D $ is \textit{the} set $ C $. So, every correct process $ P $ must decide by following the so called \textit{Trim Rule} as the rest of the correct processes in the same position. Thus, if $ D $ is \textit{the} set $ C $, the rest of processes in $ D $ will decide the same agreement value.

		The \textbf{\textit{Trim Rule}} requires every correct process to compare the received \textit{agreement proposals}, select the largest set of processes that made \textit{agreement proposals} containing the \textit{initial values} of all processes in the set, and trim every \textit{initial value} that is not in the \textit{agreement proposal} of every process in the selected set.

		At Phase Two the size of set $ C $ might have shrunk but the number of processes in $ C $ is still $ \ge (n-f) $ according to the system model. So all processes in $ C $ at the end of Phase Two decide the same agreement value, unless two sets of equal size satisfy the criteria for being \textit{the} set $ C $ and the processes in their intersection cannot decide how to apply the \textit{Trim Rule}. We will show that this cannot happen.

	\section{Formal}	
	
	
	\subsection{Tolerance to Communication Asynchrony}

		Indirect delivery uses gossip-based reliable multicast for systems where all processes are known \cite{BirmanEtAl1999}. If the link for delivery of message $m_{s}$ from process $P_{s}$ to process $P_{r}$ is faulty or asynchronous, but an indirect channel over correct synchronous links exists, $m_{s}$ will be delivered  over that channel. 

		\textbf{Lemma 1}: \textit{If only one indirect channel from sender to recipient exists, epidemic dissemination ensures delivery over this channel}.

		\textbf{Proof}

		Let $ S $ be a system that implements epidemic dissemination, every process knows every other, process $P_{s}$ has sent a message $m_{s}$ to every process in $ S $, and the link $ L_{sr} $ from $P_{s}$ to $P_{r}$ is faulty.

		The algorithm: On first receiving of $m_{s}$, every process $P_{i}$ forwards $m_{s}$ to every process that is not part of the chain that relayed $m_{s}$ from $P_{s}$ to $P_{i}$. 
		Thus, it ensures that on the existence of a single indirect channel from $P_{s}$ to $P_{r}$, message $m_{s}$ will be delivered to $P_{r}$.

		\textbf{Lemma 2}: \textit{ If a synchronous system's bound for delivery is a multiple of single-hop bound by max allowed hops, epidemic dissemination ensures highest tolerance to faulty links with no effect on synchrony}.

		\textbf{Proof}: Epidemic dissemination circumvents all faulty links that \textit{max allowed hops} makes circumventable. It follows from Lemma 1.

		\textbf{Theorem 1}: \textit{Synchronous consensus system that uses epidemic dissemination ensures bounded delivery of the same set of messages when all links of a set of links are either faulty or asynchronous.}

		\textbf{Proof}

		Let $ S $ be a system of $ n $ correct processes and epidemic dissemination, $ M $ be the smallest set of sufficient for consensus messages
		to be delivered to the smallest set of sufficient for consensus processes $ P $, and the system bound $ T $ for synchronous delivery be adjusted for the max allowed hops for indirect delivery. The links of $ S $ are: i) Case 1: Synchronous with $ F $ links faulty; or ii) Case 2: Non-faulty partially synchronous with $ A $ links operating asynchronously.

		In Case 1, $ M $ will be delivered over $ C $, the smallest sets of correct synchronous links that allows delivery of $ M $ to set $ P $ [Lemma 1] tolerating the highest number of faulty links $ F $, within the bound T [Lemma 2], where $ F = n*(n-1) - |C| $. 

		In Case 2, the same set of links $ C $ delivers the same set of messages $ M $ to the same set of processes $ P $ over the same indirect channels, thus also within the bound T, and tolerates $ A $ asynchronous links, where $ A =  n*(n-1) - |C| $. Hence $ A=F $ and in both cases system $ S $ preserves the bounded delivery of the critical set of messages $ M $.		
	
	\subsection{Consensus Protocol}

		\textit{Consensus:} Processes $P_i$, $i\in\{1,...,n\}$, start with individual initial value $v_i$ that can be a vector with finite number of elements, individually compose an $ n $-vector $V_i=(v_1,...,v_n)$, and terminate when a majority of processes $P_i$ agree on content of a vector $V=(v_1,...,v_n)$. 

		\textit{Agreement}: No less than $ (n-f) $ processes, $ n>2f $, where $ f $ is the number of tolerated crash-failed and Byzantine processes, agree on the content of an $ n $-vector.

		\textit{Validity}: The agreed-upon $ n $-vector contains as elements at least the initial values of the agreed processes and the rest of elements contain a null marker $ \varnothing $. 

		\textit{Termination}:  No less than $ (n-f) $ processes agree on an $ n $-vector within a time interval $ T_T \leq 2*3* T_S$, where $T_S$ is the longest delivery time considered synchronous.

		\textbf{Theorem 2}: \textit{A consensus protocol in partial synchrony can be correct with bounded termination in spite of possible faulty processes and faulty or asynchronous links if the faulty processes are a minority and a simple majority of correct processes operates with bounded delivery of up to three hops}.

		\textbf{Proof}

		Consensus requires every process to distribute to every other process the dataset in its $R^{In}$ and $(n-f)$ processes, where $ f $ is the number of tolerated faulty processes and $ n>2f $, to agree about 
		an ordered aggregation of individual datasets in a single dataset. 
		Sufficient is to show that every admissible run is a deciding run for no less than $(n-f)$ processes with no infringement of validity.
		The proof follows from the proofs of the following lemmas:

		\textbf{Lemma 3}: \textit{No initial configuration can prevent correct processes from reaching a decision state.}

		\textbf{Proof}

		A \textit{decision run} of a process $ P $ starts from its initial state and reaches a decision state after execution of a finite sequence of steps by $ P $ and by at least $(n-f-1)$ other processes. 
		A step may start with receiving a message and may end with sending a finite set of messages.  
		The agreement is about an ordered aggregation of data; not a decision based on that data. Hence, reaching a decision state by $ P $ depends on whether $ P $ has received an \textit{Initial Value} and \textit{Agreement Proposal} messages from at least $(n-f-1)$ processes and in no way depends on the content of the received \textit{initial values}.

		\textbf{Axiom 1}: \textit{In a consensus system of $ n $ processes, which: i) operates with bounded delivery with up to 3 hop; ii) tolerates $ f < n/2 $ process faults; and iii) there exists a set $ C $ containing $ \ge (n-f) ) $ correct process with bounded delivery within $ C $, \textbf{coexistence} of two sets, $ D $ and $ E  $, $ D \subset C  \land E \subset C \land D \ne E $, with bounded delivery within each set and $ ((|D| = |E| \ge (n-f) \land (D \cap E < (n-f))) \lor ((|D| = |E| \ge (n-f) \land (D \cap E > (n-f))) $ is \textbf{impossible}.}

\begin{table}[ht]
	\caption{Impossibility of $(|D =|E|\ge(n-f)\land(D\ne E)$}
	\begin{tabular}{c c c c c c c c c c}
		\hline 
		$ n $			& 6& 6& 6& 7& 7& 7& 7& 7& 7 \\
		$f_{tolerated}$ & 2& 2& 2& 3& 3& 3& 3& 3& 3 \\
		$n_{quorum} $	& 4& 4& 4& 4& 4& 4& 4& 4& 4 \\
		$f_{actual} $ 	& 0& 0& 1& 0& 0& 0& 1& 1& 2 \\
		$|D| = |E| $	& 5& 4& 4& 6& 5& 4& 5& 4& 4	\\
		$|D \cap E| $ 	& 4& 2& 3& 5& 3& 2& 4& 3& 3	\\	
		3-hop-Fsbl		&yes&no&yes&yes&yes&no&yes&yes&yes	\\
		Needed AL		&16& -&12&22&26& -&14&16&12	\\
		Tolertd AL		& 8& 8& 4&17&17&17& 8& 8& 5	\\
		Possibility		&no&no&no&no&no&no&no&no&no	\\	
		\hline
	\end{tabular}		
\end{table} 

		\textit{Comments}: In Table 1, the row \textit{3-hop-Fsbl} shows feasibility with 3-hop bounded delivery of a configuration with the parameters in the previous rows. Row \textit{Possibility} reflect possibility of the analysed configuration. If a configuration is not feasible with 3-hop delivery, it cannot satisfy the abstractly defined configuration.
		Row \textit{Tolertd AL} shows the maximum number of tolerated asynchronous (and/or faulty) links per number of the actual $f_{actual} $ faulty processes in a system with size $ n $. Its values are obtained experimentally and presented in the Appendix. Row \textit{Needed AL} shows the minimum of asynchronous (and/or faulty) links that make coexistence of $|D =|E|\ge(n-f)$ possible with $D \cap E < (n-f)$ or with $D \cap E > (n-f)$. The reader can verify these numbers' correctness. 
		If implementation of the analysed configuration is feasible with 3-hop bounded delivery, its possibility depends on whether the minimum asynchronous links required for its implementation can be satisfied within the maximum tolerated asynchronous links. The configuration is not possible in every case where the required minimum is higher than the tolerated maximum.

		\textbf{Lemma 4}: \textit{An algorithm can ensure within a bounded interval of time that every admissible run is a deciding run where a majority of processes decide the same valid agreement value, in the \textbf{presence} of transient asynchrony of links and \textbf{absence} of process faults.}

		\textbf{Proof}

		Links asynchrony is transient. We have to show that some links regaining their synchronous operations at any time during the consensus round cannot spoil agreement and validity.

		Let system $ S $ starts Phase One with a set $ C $ of processes with bounded delivery within $ C $, where $|C| \ge (n-f) $,  $|C| < |S| $, and  $ C $ is the largest in $ S $ set with bounded delivery within the set. No other set of processes in $ S $ with bounded delivery within the set can have the same size as $ C $ [Axiom 1].

		If no links in $ S $ have regained synchronous delivery during Phase One, or the links with regained synchronous delivery have not helped a process not in $ C $ to establish bounded delivery with every process in $ C $, system $ S $ completes Phase One with a set $ C’ \subseteq C $ of correct processes with bounded delivery within $ C' $, where $|C'| \ge (n-f) $. The bounded delivery in $ C $ ensures that $ \forall P_i \in C' \Rightarrow $ the \textit{Initial Value} message of $ P_i $ was delivered to every other process in $ C' $ and every process in $ C' $ has received an \textit{Initial Value} message from every other process in $ C' $. Existence of two versions of $ C' $ is impossible [Axiom 1]. 

		Let at a point of time after $ S $ started Phase One, one or more links in $ S $ regained synchronous delivery and, as a consequence of that, a process $ P_x $ regained its bounded delivery with every process in $ C $ and thereby established a new largest in $ S $ set $ X $ with bounded delivery within the set, so that $ X = C + P_x $, $ C \subset X $. Thus $ S $ completed Phase One with a set $  X' \subseteq X $, where $ C' \subset X' $ and $ X'-C'= P_x $. Existence of two versions of $ X' $ is impossible [Axiom 1].

		System  $ S $ starts Phase Two with sets $ X' $ and $ C’ $. If no links in $ S $ have regained synchronous delivery during Phase Two, or the links with regained synchronous delivery have not helped a
		process not in $ X' $ to establish bounded delivery with every process in $ X' $ , or a process not in $ C' $ to establish bounded delivery with every process in $ C'$, system $ S $ completes Phase Two with a set $ C’' \subseteq C' $ of correct processes with bounded delivery within $ C'' $, where $ |C''| \ge (n-f) $, and with a set $ X’' \subseteq X' $ of correct processes with bounded delivery within $ X'' $, where $ X''-C''= P_x $ if $ P_x $ has not lost its bounded delivery to or from every process in $ C'' $, or  $ X''=C'' $ otherwise.  Existence of two versions of $ C'' $ or $ X'' $ is impossible [Axiom 1].

		Let at some point of time after $ S $ started Phase Two, one or more links in $ S $ regained synchronous delivery and, as a consequence of that, a process $ P_y $ regained its bounded delivery with every process in $ X' $ and thereby established the new largest in $ S $ set $ Y $ with bounded delivery within the set, so that $ Y = X' + P_y $ and $ C' \subset X' \subset Y $.  Existence of two versions of $ Y $ is impossible [Axiom 1].
		Thus $ S $ completes Phase Two with a set $  Y' \subseteq Y $, where $ C'' \subset X'' \subset Y' $,  $ Y'-X''= P_y $ if $ P_y $ has not lost its bounded delivery to or from every process in $ X'' $, and $ X''-C''= P_x $ if $ P_x $ has not lost its bounded delivery to or from every process in $ C'' $. Existence of two versions of $ C'' $ or $ X'' $ or $ Y' $ is impossible [Axiom 1].		

		As system $ S $ operates with no faulty processes, it can reach agreement after completion of Phase Two. At the end of Phase Two,
		the bounded delivery in $ X' $ ensures that $ \forall P_j \in X'' \Rightarrow $ the \textit{Agreement Proposal} message of $ P_j $ was delivered to every process in $ X'' $ and $ P_j $ has received an \textit{agreement proposal} from every other process in $ X'' $. We have to show that every process $ P_k \in C'' $, $ (P_k \ne P_x \land P_k \ne P_y) $, decides the same value satisfying the validity conditions.

		\textit{Case 1}: Process $ P_x $ regained bounded delivery with all processes in $ C $ immediately \textbf{before} broadcasting an \textit{Initial Value} message. As a consequence, all other processes in $ X' $ received its message and $ P_x $ received an \textit{Initial Value} message from every other process in $ X' $. Thus every process in $ X' $ has broadcast an \textit{Agreement Value} message containing an initial value from at least every process in $ X' $ and, after applying the \textit{Trim Rule}, every process in $ X'' $ will decide the same valid value containing at least the \textit{initial value} of every process in $ X' $. $ \forall P_k \in C'' \Rightarrow $ $ P_k $ has received $ \ge (n-f) $ \textit{Agreement Proposal} messages, each containing the \textit{initial value} of every process in $ X' $.

		\textit{Case 2}: Process $ P_x $ regained bounded delivery with all processes in $ C $ \textbf{during} the broadcast of \textit{Initial Value} message. As a consequence, before the end of Phase One, a fraction of processes $ C'_R \subset C' $ received an \textit{Initial Value} message from $ P_x $ and all processes $  C' $ delivered their \textit{Initial Value} messages to $ P_x $.

		- \textit{Case 2A}: $ |C'_R| < (n-f) $. Consequently, $ < (n-f) $ processes in $ C' $ included the \textit{initial value} of $ P_x $ in their \textit{agreement proposals} and $ P_x $ included the \textit{initial values} of all processes in $ C' $. At the end of Phase Two, after applying the \textit{Trim Rule }: if $ |C''_R + P_x| = |C''| = (n-f) $, all processes in $ C'' $ decide agreement value containing the initial value of $ P_x $; otherwise all processes in $ C'' $ decide agreement value that does not contain the initial value of $ P_x $.

		- \textit{Case 2B}: $ |C'_R| \ge (n-f) $. Consequently, $ \ge (n-f) $ processes in $ C' $ included the \textit{initial value} of $ P_x $ in their \textit{agreement proposals} and $ P_x $ included the \textit{initial values} of all processes in $ C' $. At the end of Phase Two, after applying the \textit{Trim Rule }: if $ |C''_R + P_x| = |C''|$, all processes in $ C'' $ decide agreement value containing the initial value of $ P_x $; otherwise all processes in $ C'' $ decide agreement value that does not contain the initial value of $ P_x $.

		\textit{Case 3}: Process $ P_x $ regained bounded delivery with all processes in $ C $ \textbf{after} broadcasting an \textit{Initial Value} message. As a consequence, before the end of Phase One, a fraction of processes $ C'_R \subset C' $ received an \textit{Initial Value} message from $ P_x $ and a fraction of processes $  C'_D \subset C' $ delivered their \textit{Initial Value} messages to $ P_x $.

		- \textit{Case 3A}: $ |C'_R| < (n-f) $ and $ |C'_D| < (n-f) $. Consequently, $ < (n-f) $ processes in $ C' $ included the \textit{initial value} of $ P_x $ in their \textit{agreement proposals} and $ P_x $ included the \textit{initial values} of $ < (n-f) $ processes in $ C' $. At the end of Phase Two, $ \forall P_k \in Y' \Rightarrow $ $ P_k $ has received \textit{agreement proposals} from $ < (n-f) $ processes in $ Y' $ containing the \textit{initial value} of $ P_x $ and the \textit{Trim Rule} will enforce its replacement with $ \varnothing $ in its decided agreement value.  

		- \textit{Case 3B}: $ |C'_R| \ge (n-f) $ and $ |C'_D| < (n-f) $. Consequently, $ \ge (n-f) $ processes in $ C' $ included the \textit{initial value} of $ P_x $ in their \textit{agreement proposals} and $ P_x $ included the \textit{initial values} of $ < (n-f) $ processes in $ C' $. At the end of Phase Two, every process in $ C'' $ received from every other process in $ C'' $ an   \textit{agreement proposal} containing no less than the \textit{initial value} of every process in $ C' $. Let $ C'-C''= P_p $ and $ C'- C'_R = P_q $. Applying the \textit{Trim Rule}, causes every process in $ C'' $ to decide value containing the \textit{initial value} of $ P_x $ on condition that $P_p = P_q $; otherwise every process in $ C'' $ decides value that does not contain the \textit{initial value} of $ P_x $.

		- \textit{Case 3C}: $ |C'_R| < (n-f) $ and $ |C'_D| \ge (n-f) $. Hence, $ < (n-f) $ processes in $ C' $ included the \textit{initial value} of $ P_x $ in their \textit{agreement proposals} and $ P_x $ included the \textit{initial values} of $ \ge (n-f) $ processes in $ C' $. At the end of Phase Two, every process in $ C'' $ received from every other process in $ C'' $ an   \textit{agreement proposal} containing no less than the \textit{initial value} of every process in $ C' $. Applying the \textit{Trim Rule}, causes every process in $ C'' $ to decide value that does not contain the \textit{initial value} of $ P_x $. 

		- \textit{Case 3D}: $ |C'_R| \ge (n-f) $ and $ |C'_D| \ge (n-f) $. Hence, $ \ge (n-f) $ processes in $ C' $ included the \textit{initial value} of $ P_x $ in their \textit{agreement proposals} and $ P_x $ included the \textit{initial values} of $ \ge (n-f) $ processes in $ C' $. At the end of Phase Two, every process in $ C'' $ received from every other process in $ C'' $ an   \textit{agreement proposal} containing no less than the \textit{initial value} of every process in $ C' $ and also all processes in $ C'' $ received $ \ge (n-f) $ \textit{agreements proposals} from other processes in $ C'' $ containing the \textit{initial value} of $ P_x $.
		Let $ C'-C''= P_p $ and $ C'- C'_R = P_q $. Applying the \textit{Trim Rule}, causes every process in $ C'' $ to decide value containing the \textit{initial value} of $ P_x $ on condition that $P_p = P_q $; otherwise every process in $ C'' $ decides value that does not contain the \textit{initial value} of $ P_x $.

		\textbf{Axiom 2}: \textit{In a consensus system of $ n $ processes, which: i) operates with bounded delivery with up to 3 hop; ii) tolerates $ f < n/2 $ process faults; and iii) there exists a set $ C $ containing $ \ge (n-f) ) $ correct process with bounded delivery within $ C $, \textbf{coexistence} of two sets, $ D $ and $ E  $, $ D \subset C  \land E \subset C \land D \ne E $, with bounded delivery within each set and 
		$ ( |E| > |D| \ge (n-f) \land (D \cap E < (n-f) ) $ is \textbf{impossible}.}

\begin{table}[ht]
	\caption{Impossibility of $(|E|>|D|\ge (n-f)\land (D \cap E<(n-f)) $}
	\begin{tabular}{c c c c c c c c c c}
		\hline 
		$ n $			& 5& 6& 6& 7& 7& 7& 8& 8& 8 \\
		$f_{tolerated}$ & 2& 2& 2& 3& 3& 3& 3& 3& 3 \\
		$n_{quorum} $	& 3& 4& 4& 4& 4& 4& 5& 5& 5 \\
		$f_{actual} $ 	& 0& 0& 1& 0& 0& 1& 0& 0& 1 \\
		$|D| $			& 3& 4& 3& 4& 3& 4& 5& 4& 5	\\
		$|E| $			& 4& 5& 4& 5& 4& 5& 6& 5& 6	\\
		$|D \cap E| $ 	& 2& 3& 2& 3& 2& 3& 4& 3& 4	\\	
		3-hop-Fsbl		&no&yes&no&yes&no&yes&yes&yes&yes	\\
		Needed AL		& -&16& -&24& -&18&38&44&24	\\
		Tolertd AL		& 9& 8& 5&17&17& 8&15&15&11	\\
		Possibility		&no&no&no&no&no&no&no&no&no	\\	
		\hline
	\end{tabular}		
\end{table} 	

		\textit{Comments}: In Table 2, the row \textit{3-hop-Fsbl} shows feasibility with 3-hop bounded delivery of a configuration with the parameters in the previous rows. Row \textit{Possibility} reflect possibility of the analysed configuration. If a configuration is not feasible with 3-hop delivery, it cannot satisfy the abstractly defined configuration.
		Row \textit{Tolertd AL} shows the maximum number of tolerated asynchronous (and/or faulty) links per number of the actual $f_{actual} $ faulty processes in a system with size $ n $. Its values are obtained experimentally and presented in the Appendix. Row \textit{Needed AL} shows
		the minimum of asynchronous (and/or faulty) links that make coexistence of $|E|>|D|\ge (n-f) $ with $(D \cap E<(n-f) $ possible. The reader can verify numbers' correctness. If the analysed configuration is feasible with 3-hop bounded delivery, its possibility requires the minimum asynchronous links needed for its implementation to be within the maximum tolerated asynchronous links.	

		\textbf{Lemma 5}: \textit{An algorithm can ensure within a bounded interval of time that every admissible run is a deciding run where a majority of processes decide the same valid agreement value, in the \textbf{presence} of crashing processes and \textbf{absence} of transient asynchrony of links.} 

		\textbf{Proof}	

		We have to demonstrate that a consensus system $ S $ of $ n $ processes that tolerates up to $ f $ crash-fails, can tolerate $ x \le f $ crash-fails happening at any point of time of consensus round on condition that before and after every crash-fail there exists a majority of correct processes with bounded delivery between themselves, so that $ \ge (n-f) $ processes will decide the same agreement value. 

		Let system $ S $ starts Phase One with a set $ C $ of processes with bounded delivery within $ C $, where $|C| \ge (n-f) $, and completes Phase One with a set $ C’ \subseteq C $ of correct processes with bounded delivery within $ C' $, where $|C'| \ge (n-f) $. The bounded delivery in $ C $ ensures that $ \forall P_i \in C' \Rightarrow $ the \textit{Initial Value} message of $ P_i $ was delivered to every correct process in $ C $ and every process in $ C' $ has received an \textit{Initial Value} message from every other process in $ C' $. Existence of two versions of $ C' $ is impossible according to Axiom 1.

		System  $ S $ starts Phase Two with set $ C’ $ and completes Phase Two with a set $ C’' \subseteq C' $ of correct processes with bounded delivery within $ C'' $, where $ |C''| \ge (n-f) $. The bounded delivery in $ C' $ ensures that $ \forall P_j \in C'' \Rightarrow $ the \textit{Agreement Proposal} message of $ P_j $ was delivered to every process in $ C'' $, and every process in $ C'' $ received from every other process in $ C'' $ an \textit{agreement proposal} containing no less than the \textit{initial values} of all processes in $ C'' $. 
		Two versions of $ C'' $ are impossible according to Axiom 1.

		System $ S $ decides the agreement value at the end of Phase Two.  Every process $ P_j \in C'' $ produces a \textit{decision value} by performing the following sequence of operations. It applies the \textit{Trim Rule} on every \textit{agreement proposal} received from a process in $ C'' $. The objective is to  get rid of every \textit{initial value} that is not in every \textit{agreement proposal} received from a process in $ C'' $. The trim produces $ \ge (n-f) $ vectors with the same value containing the \textit{initial value} of every process in a set $ G $, where $ C'' \subseteq G $. Process $ P_j $ can confidently use this value as decision value, except in the following cases:

		- \textit{Case 1}: Phase Two completed with two sets, $ D $ and $ E $, of equal size satisfying the criteria for being \textit{the} set $ C'' $ and every process $ P_j \in D \cap E $ cannot decide how to apply the \textit{Trim Rule} without knowing the identities of the processes that crash-failed after broadcasting their agreement proposals. This case is not feasible as the presented configuration is not feasible [Axiom 1].

		- \textit{Case 2}: Phase Two completed with two set $ D $ and $ E $, where $ |D| = (n-f) \land |E| = (|D|+1) \land D \cap E = (n-f-1) ) $. The \textit{Trim Rule} requires and every process $ P_j \in D \cap E $ to select set E. However, if processes $ P_q: P_q \in E \land P_q \notin D $ and $ P_r: P_r \in E \land P_r \notin D $ both crash-failed after broadcasting their \textit{Agreement Proposal} messages, system $ S $ would have been within the system model assumptions but an the end of consensus round $ < (n-f) $ would decided the same agreement value. This case is not feasible as the presented configuration is not feasible according to Axiom 2.

		At the end of consensus round, $ C'' \subseteq C' \subseteq C $, The processes in $ C'' $ do not know how many among them might have crashed during the broadcast of \textit{Agreement Proposal} message and how many after. Yet every process in $ C'' $ knows that: i) $ |C''| \ge (n-f) $ according to the system model; and ii) all processes in $ C'' $ decide the same value.

		\textbf{Lemma 6}: \textit{An algorithm can ensure within a bounded interval of time that every admissible run is a deciding run where a majority of processes decide the same valid agreement value, in the \textbf{presence} of Byzantine processes and \textbf{absence} of transient asynchrony of links.} 

		\textbf{Proof}		

		Byzantine processes can omit to send to all recipients, delay before sending to one or more, and deceit in regard to message content. We have to demonstrate that Byzantine processes cannot do more harm than crash-failed processes. 

		\textit{Case 1}: Omission to send its own protocol message, or to retransmit a received protocol message as part of an indirect delivery channel.

		- \textit{Case 1A}: Omission to send own protocol message to all recipients. The effect is the same as with a process crash-fail before the broadcast. It cannot spoil anything [Lemma 5].

		- \textit{Case 1B}: Omission to send own protocol message to some recipients. The effect is the same as with a process crash-fail during the broadcast. It cannot spoil anything [Lemma 5].

		- \textit{Case 1C}: Omission to retransmit a received protocol message as part of an indirect delivery channel to all recipients or to some recipients. The effect is the same as with a process crash-fail after the broadcast. It cannot spoil anything [Lemma 5]. 

		\textit{Case 2}: Delay before sending own protocol message to some or all recipients, or before retransmission of a received protocol message as part of an indirect delivery channel. 

		- \textit{Case 2A}: Delay before sending own protocol message to some or all recipients and before retransmission of a received protocol message as part of an indirect delivery channel to some or all recipients. The effect is the same as with a correct process with asynchronous outgoing links before the broadcast and after the broadcast. It cannot spoil anything [Lemma 4].

		- \textit{Case 2B}: Delay before sending own protocol message to some or all recipients but not before retransmission of a received protocol message as part of an indirect delivery channel to some or all recipients. The effect is the same as with a correct process operating with asynchronous outgoing links that existed before the broadcast but regained synchrony after the broadcast. It cannot spoil anything [Lemma 4].

		- \textit{Case 2C}: Delay before retransmission of a received protocol message as part of an indirect delivery channel to some or all recipients. The effect is the same as with a correct process operating with asynchronous outgoing links that appeared  before retransmission of a received protocol message. It cannot spoil anything [Lemma 4].

		\textit{Case 3}: Deceit in regard to message content. A Byzantine process cannot send signed \textit{Initial Value} messages with different content without revealing itself, but it can omit sending its own \textit{Initial Value} message to some processes. It cannot include in its \textit{Agreement Proposal} message \textit{initial values} it has not received, but it but can omit including a received \textit{initial values} in its \textit{Agreement Proposal} message. Thus the following attacks can be attempted without Byzantine processes being revealed:

		- \textit{Case 3A}: The set $ C $ contains $ (n-f) $ correct processes and all processes in $ C $ consider a Byzantine process $ P_B $ as being a correct one in $ C $. Process $ P_B $ omits to send its \textit{Initial Value} message to a process $ P_D \in C $ and omits to include the \textit{initial value} of $ P_D $ in its own agreement proposal. At the end of Phase Two, $ (n-f-1) $ processes in $ C'' $ have to apply the \textit{Trim Rule} over $ (n-f-1) $ \textit{agreement proposals} containing the \textit{initial values} of both $ P_B $ and $ P_D $. Assuming the existence of a way to make those $ (n-f-1) $ processes trim the same \textit{initial value}, either the \textit{initial value} of $ P_B $ or the \textit{initial value} of $ P_D $, and thereby ensure $ (n-f) $ processes making the same decision, there is a probability of 0.5 that the trimmed \textit{initial value} is the one of the correct process $ P_D $. Consequently, less than $ (n-f) $ correct processes would have decided the same agreement value even though the system contains $ (n-f) $ correct processes with bounded delivery between themselves. This case is not feasible according to Axiom 1.

		- \textit{Case 3B}: The set $ C $ contains $ (n-f) $ correct processes and all processes in $ C $ consider Byzantine processes $ P_{B1} $ and $ P_{B2} $ as being correct ones in $ C $.  Processes $ P_{B1} $ and $ P_{B2} $ omit to send their \textit{Initial Value} messages to a process $ P_D \in C $ and omit to include the \textit{initial value} of $ P_D $ in their agreement proposal.  At the end of Phase Two, $ (n-f-1) $ processes in $ C'' $ are required by the \textit{Trim Rule} to trim the value of $ P_D $. Consequently, less than $ (n-f) $ correct processes would have decided the same agreement value even though the system contains $ (n-f) $ correct processes with bounded delivery between themselves. This case is not feasible according to Axiom 2.

		\textbf{Lemma 7}: \textit{An algorithm can ensure within a bounded interval of time that every admissible run is a deciding run where a majority of processes decide the same valid agreement value, in the \textbf{presence} of crashing and Byzantine processes, and transient asynchrony of links.} 

		We have to demonstrate that during a consensus round some links regaining synchrony, together with the crashed and Byzantine processes, cannot prevent agreement or its validity.

		Let $ n_F $ be the number of processes that crashed before the start of current consensus round, $ n_B $ be the number of Byzantine processes, and $ n_{Cr} $ be the number of processes that may crash during the current consensus round, so that $ n_F + n_B + n_{Cr} \le f $. Trivial are the cases
		where the system model does not allow more crash-failed or Byzantine processes and new asynchronous links cannot affect $ |C|$: i) i) $ n_F = f $; and ii) $ (n_F < f) \land ( |C|= (f+1)) $. \textit{Non-trivial configuration} of a consensus system is:
		$ (( n_F + n_B + n_{Cr} ) < f) \land (|C| > (f+1)) \land (( n_F + n_B + n_{Cr} +|C|) < n ) $.  It allows more processes in $ C $ to crash or become Byzantine, more links to become asynchronous, and more correct processes to establish two-way bounded delivery with every process in $ C $ from asynchronous links regaining synchrony. The system of Lemma 4 fits into the \textit{non-trivial configuration}, except in regard to possibility for crash-failed and Byzantine processes. We have to show that adding to Lemma 4 the possibility for crash-failed and Byzantine processes cannot affect its correctness.
		
		Let $ P_{Cr} $ be a process that may crash during consensus round and $ P_A $ be a process that may drop from the largest set of processes with bounded delivery due to asynchrony of links. Considering the possible dynamic disappearing of asynchrony, Lemma 4 operates with sets $ C’' \subseteq C’ \subseteq C $ of correct processes with bounded delivery within $ C'' $, where $|C''| \ge (n-f) $ according to the system model. $ C’ \subseteq C $ reflects a possibility that processes $ P_{Cr} $ and $ P_A $ might have been in $ C $, but were not in $ C' $ at the end of Phase One.
		$ C’' \subseteq C' $ reflects a possibility that processes $ P_{Cr} $ and $ P_A $ might have been in $ C' $, but were not in $ C'' $ at the end of Phase Two. 
		When $ |C|=|C'|+2 $ and $ |C'|=|C''|+2 $, Lemma 4 considers that dynamically increased number of asynchronous links caused the difference. However, if it is caused by $ P_A $ and $ P_{Cr} $, this cannot affect the correctness of Lemma 4. 
		Moreover, when $ |C|=|C'|+3 $ and $ |C'|=|C''|+3 $, in view of Lemma 6 about the inability of a Byzantine process $ P_B $ to cause more harm than $ P_{Cr} $, when the difference is caused by $ P_A $, $ P_{Cr} $, and $ P_B $, the correctness of Lemma 4 would be still unaffected.	


	\section{Summary of Findings}


		In pursuing the objectives of this work using 3-hop indirect delivery, we found:
		
		1. The number of tolerated link faults / links asynchrony per system size and the actual number of faulty processes.
		
		2. Impossibility of two equal in size largest consensus majority sets operating with bounded delivery.
		
		3. Impossibility of two consensus majority sets of different size with bounded delivery and intersection smaller than a majority.
		
		4. Possibility for $ f $-resilience to Byzantine faults implemented with two rounds of exchange of messages.
		
		5. Impossibility of transient nature of links asynchrony and link faults to affect consensus safety, validity, or termination.
		
		6. The same tolerance to max allowed process crashes regardless of whether crashes happened before or during consensus rounds.
		
		7. A way to make leaderless consensus in partial synchrony a non-existing theoretical or practical problem.
		
		8. Possibility for consensus algorithm in partial synchrony with
	 	ensured simultaneous validity, safety, and bounded liveness.
	
		
	\section{Conclusion}	
		
		
		We proved the possibility for deterministic Byzantine consensus with synchronous termination in partial synchrony, by showing: i) possibility to tolerate number of asynchronous links, that is a function of the number of faulty processes per system size, with the use of 3-hop epidemic dissemination; and ii) an algorithm that operates with cryptographically authenticated messages and ensures resilience to $ f $ Byzantine processes with 2 messaging rounds. 

		A round with circumvented asynchrony involves exchange of $ n^3 $ messages and lasts 3 times longer than a round in synchrony, where $ (f+1) $ rounds are needed for resilience to $ f $ Byzantine processes \cite{DolevStrong1983}. Yet a system, resilient to more than 5 Byzantine processes, terminates with the presented solution in partial synchrony faster than with the best known Byzantine algorithm in synchrony. This is paid with 13.56 percent more exchange of data.\footnote{
		With 100kB initial data per process, 32 bytes hash digest size, and 64 bytes signature size, a system of 11 processes tolerates 5 Byzantine processes, requires 6 communication rounds in synchrony, exchanges 86.12 MB of data with 880 messages, and terminates after 6 times the synchrony delivery bound \cite{DolevStrong1983}. With our algorithm, the same system tolerates the same number of Byzantine processes and terminates within the same time interval in partial synchrony with exchange of 1,980 messages. With the same workload, hash digest size, and signature size, it exchanges 97.80MB of data – just 13.56 percent more.}

		The presented solution enables the design of non-blocking distributed transactions across decentralised database systems. These transactions are vital for the feasibility of some world-shaking innovations. The International Monetary Fund concept about foreign exchange markets with tokenised currencies and instant settlement \cite{TobiasAdrian_EtAl_2023} showcases their critical role.

\begin{acks}
	This work is partly sponsored by the Australian Federal Government through the Research and Development Tax Incentive Scheme. 

\end{acks}


	\bibliographystyle{ACM-Reference-Format}
	\bibliography{SCIPS}
	
	\clearpage
	
	\section*{Appendix}

	\vspace{\baselineskip}

\section{Boundaries of Tolerance}

For a system with a particular number of total processes we need to compute all permutations of faulty / asynchronous links and faulty (crash-failed or Byzantine) processes per each tolerated number of faulty processes. For every computed permutation we must test whether each process in a group of $(F+1)$ correct processes receives a message from every process in that group. This is the necessary and sufficient condition for solving consensus. The objective is to discover the highest number of faulty and/or asynchronous links per number of crash-failed and/or Byzantine processes where every permutation solves consensus.

\subsection{The Simulator}

Individual permutations of the same number of correct synchronous links may need a different number of 2-hop and 3-hop indirect channels to solve consensus. A system of 7 processes requires analysing billions of permutations. Simulation of asynchrony with 100ms per consensus round and 1 billion consensus rounds would have taken more than 7 years uninterrupted work to complete. It is not only unrealistic but also meaningless as the results obtained with circumvention of faulty links in complete synchrony operate with the same logic as circumvention of asynchronous links in partial synchrony. Theorem 1 demonstrated tolerance to an equal number of faulty links and asynchronous links in any combination. 

Lemma 5 demonstrated the impossibility of transient nature of links asynchrony and link faults to affect consensus safety, validity, or termination. Hence, separate simulation of all combinations of link faults and link recoveries at different points of consensus rounds is not necessary. Sufficient for our purposes is to explore the possibility for consensus with a variety of faulty links at the start of consensus round.
Lemma 7 demonstrated that Byzantine processes cannot cause more harm than the same number of processes crash-failing in the same consensus round. Lemma 6 demonstrated that a system that tolerates the max allowed crash-fails before the start of a consensus round also tolerates the same number of crash-fails during the consensus round. Hence, separate simulation of the entire variety of feasible Byzantine faults and all combinations of crash-fails at different points of consensus rounds is not necessary. Sufficient for our purposes is the allowed variety of tolerated numbers of crash-fails before the start of consensus round.

Discovering the highest number of tolerated faulty links per number of crash-failed processes with solvable consensus requires a simulator to compute all permutations and test each one individually. The computation could happen on a single computer. A one-way link could be implemented as a shared memory, which can be modified by exactly one process and read by exactly one other process. With this objective, we built a simulator that 'broadcasts' the algorithm's messages according to the permutation of correct processes and correct links. As all link faults and process crash-fails happened before the start of consensus round, receiving a  message by a recipient from a sender is considered receiving of both Phase One and Phase Two messages. Thus, a completed distribution of messages is considered completion of a consensus round. It is followed by analysis for the possibility of valid agreement. 

\begin{figure}[h]
	\scalebox{0.10}
	{\includegraphics{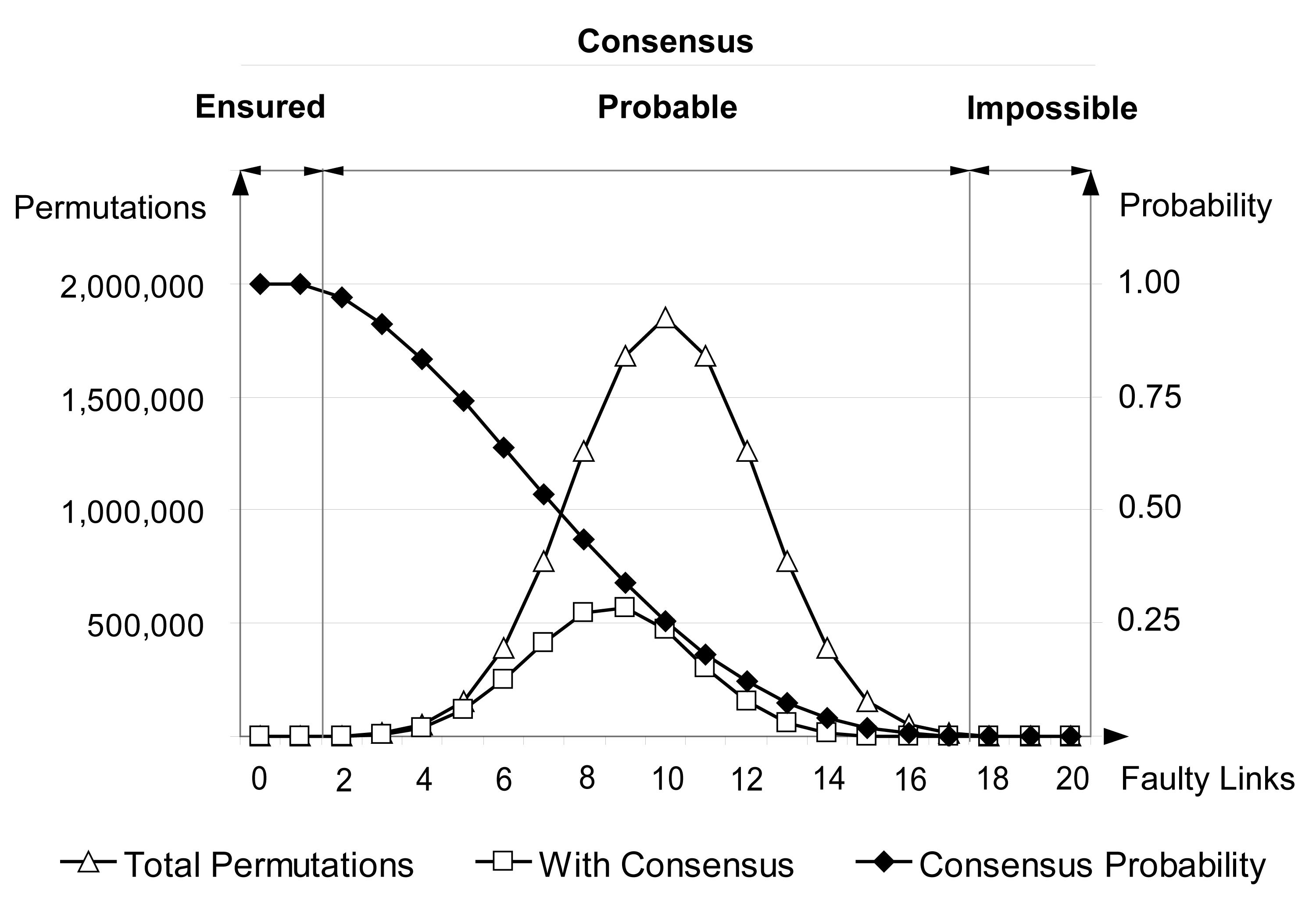}}
	\caption{Consensus with 2 Faulty of 5 Processes}
\end{figure}

The simulator operates in the following manner:

\textbf{Input}: 

It takes as input: a) total number of processes; b) number of correct processes; and c) number of correct links.

\textbf{Execution}: 

It creates objects representing processes and links, traces all permutations of correct processes and correct links with execution of the following steps:

- Simulates broadcasting of messages on behalf of every correct process via its every correct outgoing link. 

- Checks for existence of a group of $(F+1)$ correct processes, wherein every process of the group has received a message from every other process of that group. If yes, the consensus is considered solved.

\textbf{Output}: 

It returns as output: a) total number of permutations; and b) number of permutations allowing to solve consensus.

\subsection{Computation with a 5-process System}

According to the model, a 5-process system tolerates up to 2 faulty, Byzantine or crash-failed, processes. We performed three series of computations, each comprising a series of 21 computations itself – the first with 0 faulty links and the last with 20. Every computation traces all permutations of faulty processes and faulty links and tests each permutation for ability to solve consensus. 


\textbf{Series with Two Faulty Processes}

Figure 1 depicts 3 diagrams presenting:

- Total permutations, per number of faulty links;

- Permutations that solve consensus, per number of faulty links; 

- Probability to solve consensus, per number of faulty links. 

In regard to probability (number of permutations that solve consensus divided by total permutations), Figure 1 shows 3 regions:

- Region with \textbf{ensured} consensus, where probability is 1;

- Region with \textbf{probable} consensus, where $0 < $ probability $< 1$; 

- Region with \textbf{impossible} consensus, where probability is 0. 	

As it is shown on Figure 1, a 5-process system operating with 2 faulty processes ensures consensus with up to 1 faulty link. With 2 faulty links, the system solves consensus with 1,840 permutations out of 1,900 total, hence operates with 0.968 probability to solve it.


\begin{figure}[h]
	\scalebox{0.10}
	{\includegraphics{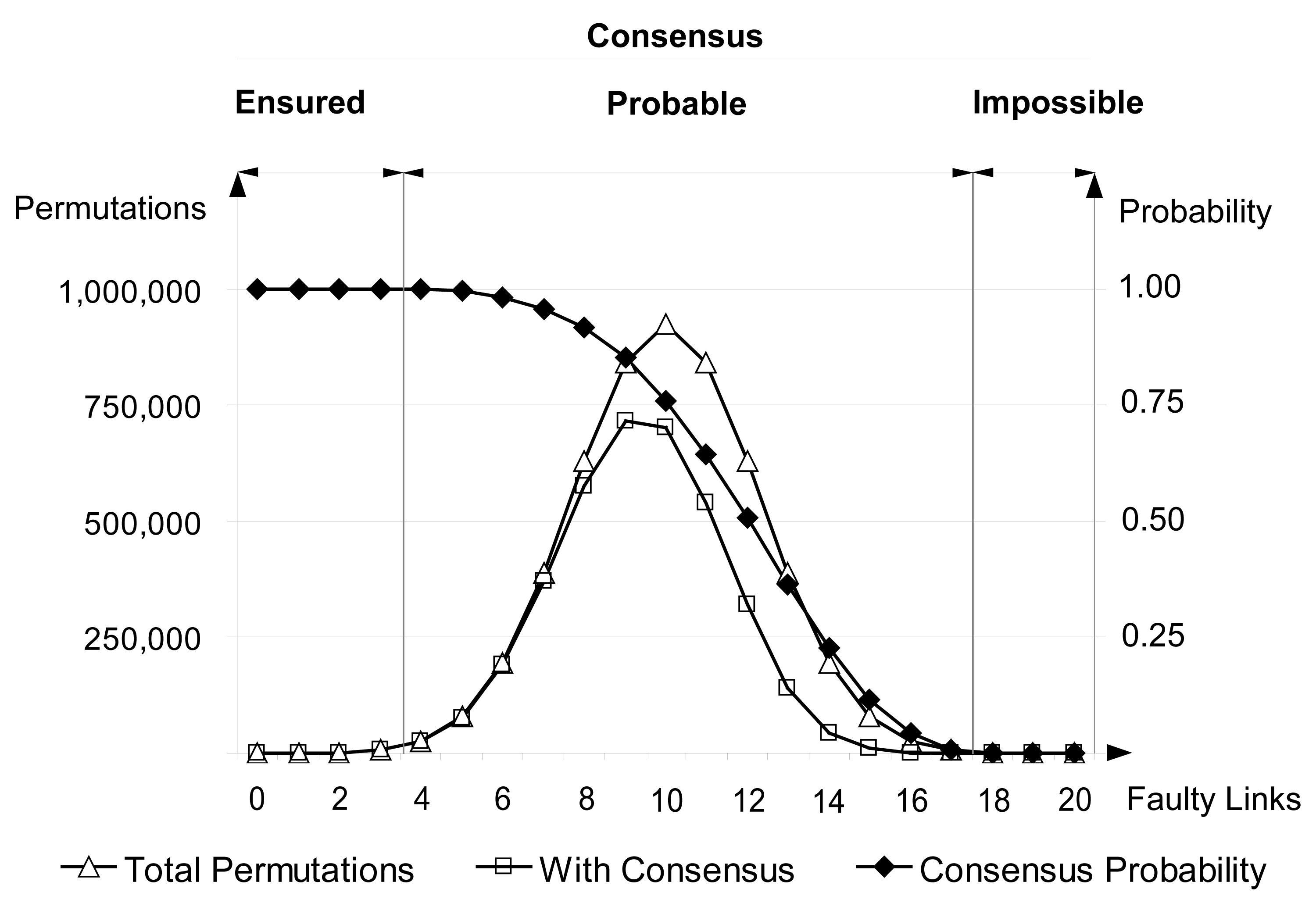}}
	\caption{Consensus with 1 Faulty of 5 Processes}
\end{figure}	

\textbf{Series with One Faulty Process}

Figure 2 also depicts 3 diagrams presenting: total permutations, permutations that solve consensus, and probability to solve consensus, per number of faulty links. 

In regard to the probability to solve consensus, the major change is expansion of the region where consensus solving is ensured. This expansion has a simple explanation with the modelling of a faulty process. All inbound and outbound links are dead with a crash-failed process or are considered dead as a worst-case scenario with a Byzantine process.

As Figure 2 shows, a 5-process system with 1 faulty process:

- Ensures consensus solving with up to 3 faulty links. 

- In presence of 4 faulty links, the system solves consensus with 24,195 permutations out of 24,225 total, hence operates with 0.9987 probability to solve it. 

- In presence of 5 faulty links, the system solves consensus with 24,074 permutations out of 24,225 total, hence operates with 0.9938 probability to solve it.

 \vspace{\baselineskip}

\textbf{Series with No Faulty Process}

Figure 3 as well depicts the diagrams presenting: total permutations, permutations that solve consensus, and probability to solve consensus, per number of faulty links. 

In regard to the probability to solve consensus, compared to the diagrams on the previous two figures, the major change again is the expansion of the region with ensured solving of consensus. Yet on this figure the expansion is huge: from tolerance of 3 faulty links with 1 faulty process to tolerance of 9 faulty links with no faulty processes.

As Figure 3 shows, a 5-process system operating with no faulty processes:

- Ensures consensus solving with up to 9 faulty links. 

- In presence of 10 faulty links, the system solves consensus with 184,696 permutations out of 184,756 total, hence operates with 0.9986 probability to solve it. 

- In presence of 11 faulty links, the system solves consensus with 167,420 permutations out of 167,960 total, hence operates with 0.9967 probability to solve it. 

\begin{figure}[h]
	\scalebox{0.10}
	{\includegraphics{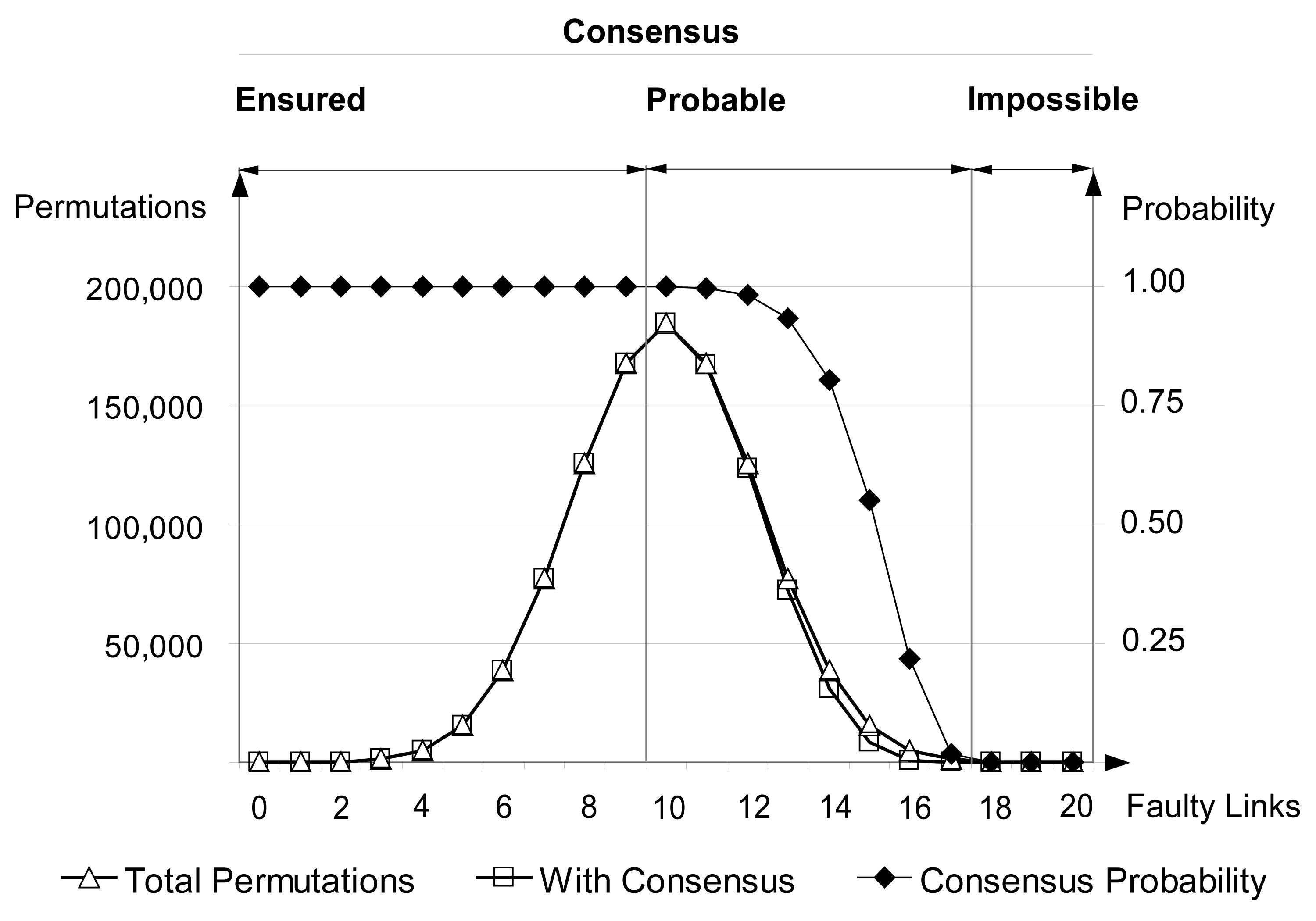}}
	\caption{Consensus with 0 Faulty of 5 Processes}
\end{figure}

\section{Handling Simulation Challenges}

Checking whether a consensus system of 9 processes can unconditionally tolerate simultaneous 3 faulty processes and 7 faulty links requires tracing and verifying for solvability of consensus 123 billion (more accurately 123,741,215,136) permutations, which took 58 days to complete the computation. The same system when operates with 2 faulty processes is expected to tolerate 11 faulty links. Checking for unconditional solvability of consensus requires tracing and verifying 108 trillion (more accurately 108,802,275,708,672) permutations, which requires 54,401 days, which is 149 years. Even after taking into consideration the symmetry factor of 9, the computation would have taken 16 years performed on a single computer. 

Hence the ability of our simulation system to compute all permutations involving faulty processes and links, and to examine each one for ability to solve consensus is practically useless with a 9-process system or larger. 
Figure 4 presents the tolerance to faulty links of a 5-process, a 7-process, and a 9-process system, computed on multiple simulation systems within a time frame of 2 months.

\subsection{The Handling Idea}

With a few exceptions, tolerance boundaries of a system of any size cannot be established with a mainstream computing technology. The alternative is to do it analytically in the following manner:

- Extract as much data as possible with computation;

- Use the computation data to fill a regression table;

- Use the regression table to build tolerance equations;

- Apply the equations to a system of any size.

Computations with a 6-process system with no faulty processes show that the system tolerates up to 4 faulty links. This means that under any permutation of 4 faulty links, the system unconditionally delivers all sent messages. The same system increases its tolerance to faulty links from 4 to 7 when it has to deliver messages of any 5 processes to the same 5 processes, and from 7 to 15 faulty links when it has to deliver from any 4 processes to the same 4 processes.

\begin{figure}[h]
	\scalebox{0.10}
	{\includegraphics{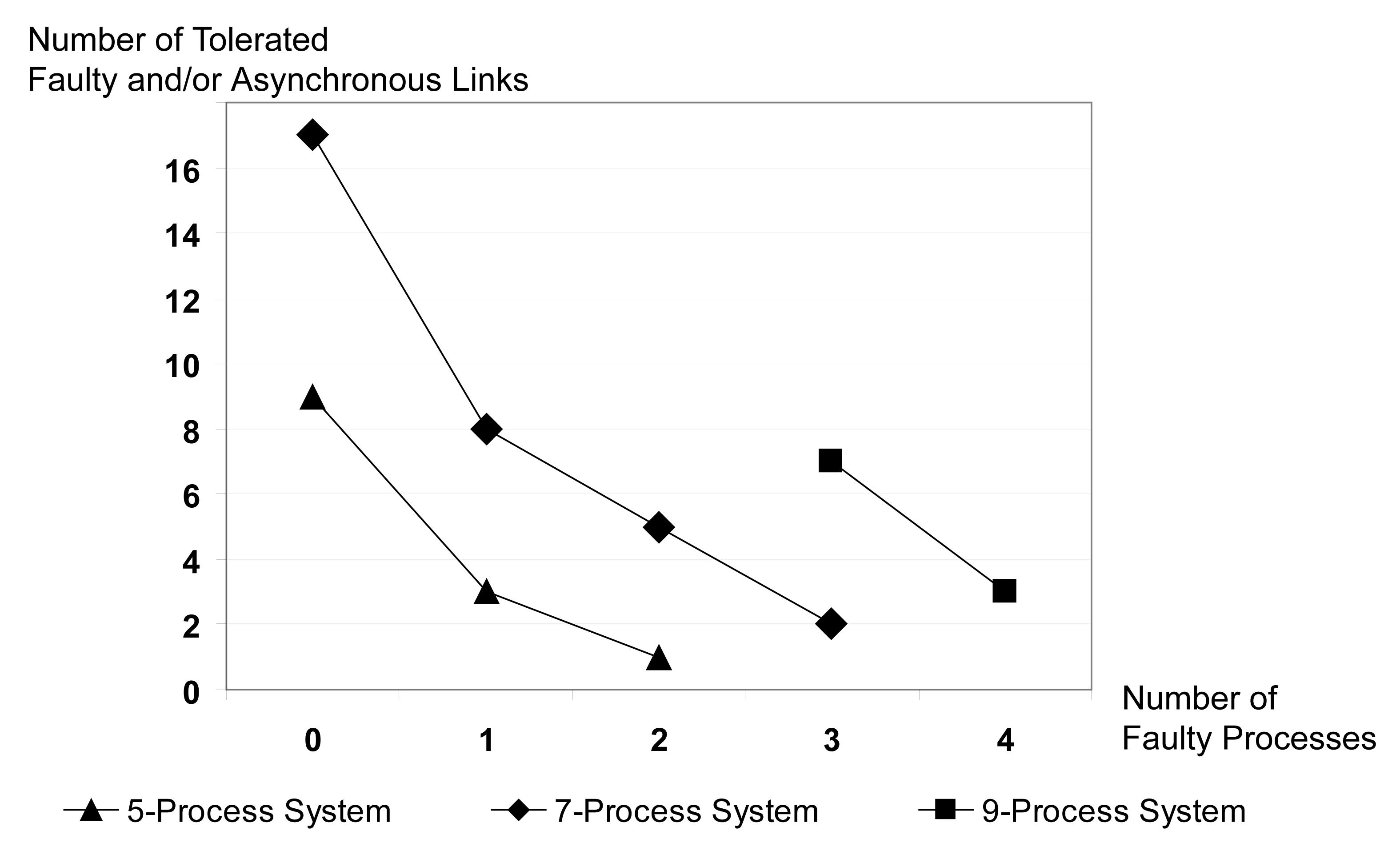}}
	\caption{Tolerance to Async/Faults (Computation Results) }
\end{figure}	

\textbf{Creation of regression table}

First step is to fill a row with the computed results showing how the tolerance to faulty links in delivery from $N$ to $N$ processes in an $N$-process system increases with an increment of $N$ with 1. Then to fill the next row with experimentally computed results showing how the tolerance to faulty links in delivery from $(N-1)$ to $(N-1)$ processes in an $N$-process system increases with an increment of $N$ with 1, then from $(N-2)$ to $(N-2)$ processes, etc. 

\textbf{Illustration of the approach}

Consider the tolerance to faulty links $f$ of the 7-process system on Figure 4 and the possible regression. The system tolerates up to $F=3$ faulty processes and number of $f$ faulty links, where with:

- $F=3$ faulty processes, it is practically a 4-process system, which tolerates $f=3$ in delivering across the 4 correct processes.

- $F=2$ faulty processes, it is practically a 5-process system, which tolerates $f=3$ in delivering across the 5 correct processes, and tolerates additional $+f=2$ faulty links in delivering across any 4 correct processes, so the resulting tolerance is $f=5$ faulty links.

- $F=1$ faulty processes, it is practically a 6-process system, whose tolerance is equal to the sum of tolerances in delivering across the 6 correct processes and additional tolerance in delivering across any 5 correct processes and across any 4 correct processes.

\subsection{Regression Table From Computations}

Table 3 presents the results of computations on multiple simulator systems within a time limit set to 2 months per computation series. Individual computation series were performed with a varied total number of processes  $N = 3, 4, 5, ..., 9$. Each one established tolerance to faulty links $f$ in delivery of $N$ messages across $N$ processes, $(N-1)$ messages across $(N-1)$ processes, etc.

Table 3 data allows observation of the following 4 patterns:

\textbf{Pattern 1}. In delivery of $N$ messages across $N$ processes, an increment of $N$ with 1 increases the faulty link tolerance $f$ with 1.

\textbf{Pattern 2}. In delivery of $(N-1)$ messages across $(N-1)$ processes, starting from $N=4$, an increment of $N$ with 1 increments $\Delta$ with 1. The meaning of $\Delta$ is the additional tolerance ($+f$), caused by simplifying the objective by 1, i.e. from delivery of $N$ messages across $N$ processes to delivery of $(N-1)$ across $(N-1)$ processes. 

\textbf{Pattern 3}. In delivery of $(N-2)$ messages across $(N-2)$ processes, starting from $N=6$ there is a repetition of Pattern 2. Hence, Pattern 3 can be formulated as: Starting from $N=4$, an increment of $N$ with 2 causes a new start of Pattern 2 in regard to a simplified by 1 objective. i.e. from delivery of $(N-1)$ messages across $(N-1)$ processes to delivery of $(N-2)$ messages across $(N-2)$ processes. 	

\begin{table}[ht]
	\caption{Computation Results}
	\begin{tabular}{c c c c c c c c c}
		\hline
		Delivery Type &N 		&3&4&5& 6& 7& 8& 9 	\\
		\hline
		N to N 	  &$f$ 		&1&2&3& 4& 5& 6& 7 	\\
		(N-1) to (N-1)&$\Delta$ &2&1&2& 3& 4& 5& 	\\
		(N-1) to (N-1)&$f$	 	&3&3&5& 7& 9&11& 	\\
		(N-2) to (N-2)&$\Delta$ & &5&4& 1& 2&  &	\\	
		(N-2) to (N-2)&$f$		& &8&9& 8&11&  &	\\
		(N-3) to (N-3)&$\Delta $& & & & 7& 6&  &	\\
		(N-3) to (N-3)&$f$		& & & &15&17&  &	\\
		\hline	\end{tabular}		
\end{table} 		

\textbf{Pattern 4}. With $N-F=F+1$, where $F$ is the highest tolerated number of faulty processes, starting from $N=3$, an increment of $N$ with 2 causes an increment of $\Delta$ with 2. This pattern is clearly observed in the following sequence: 1)) N=3, delivery from $(N-1)$ to $(N-1)$, $\Delta=2$; 2) N=5, delivery from $(N-2)$ to $(N-2)$, $\Delta=4$; and 3) N=7, delivery from $(N-3)$ to $(N-3)$, $\Delta=6$.

\subsection{Regression Table Analytically Expanded}	

Table 4 presents together the results from computation and from analytical extrapolation, based on the observed cause and effect patterns between the incremented number of system processes $N$ and the various $\Delta$s. Recall that
value of $f$ in a particular row and column is computed by adding the value of $\Delta$ in the previous row of the same column to the of preceding $f$ in that column.

\subsection{Examples: Use of Extrapolation}

Tolerance of a 9-process system to faulty links per actual number of faulty processes $F$ can be computed (with $F=4$, $F=3$, and $F=2$) using data from Table 3 and (with $F=1$ and $F=0$) using data from Table 4 in the following manner:

- With $F=4$, a 9-process system is equivalent to a 5-process system, which tolerates $f=3$ in delivering 5 messages to 5 processes (Table 3).

- With $F=3$, the system is equivalent to a 6-process system, which tolerates $f=4$ in delivering 6 messages to 6 processes (Table 1), and tolerates additional $+f=3$ in delivering 5 message to 5 processes (Table 3), so the resulting tolerance is $f=7$.

- With $F=2$, the system is equivalent to a 7-process system, which tolerates $f=5$ in delivering 7 messages to 7 process (Table 3), tolerates additional $+f=4$ in delivering 6 messages to 6 processes (Table 3), and tolerates $+f=2$ in delivering 5 messages to 5 processes (Table 3), so the resulting tolerance is $f=11$.

\begin{table}[ht]
	\caption{Computation and Extrapolation Results}
	\begin{tabular}{c c c c c c c c c c c}
		\hline
		Delivery Type &N 		&3&4&5& 6& 7& 8& 9&10&11 	\\
		\hline
		N to N 	   &$f$ 	&1&2&3& 4& 5& 6& 7& 8& 9 	\\
		(N-1) to (N-1)&$\Delta$&2&1&2& 3& 4& 5& 6& 7& 8 	\\
		(N-1) to (N-1)&$f$	 	&3&3&5& 7& 9&11&13&15&17 	\\
		(N-2) to (N-2)&$\Delta$& &5&4& 1& 2& 3& 4& 5& 6	\\	
		(N-2) to (N-2)&$f$		& &8&9& 8&11&14&17&20&23	\\
		(N-3) to (N-3)&$\Delta$& & & & 7& 6& 1& 2& 3& 4	\\
		(N-3) to (N-3)&$f$		& & & &15&17&15&19&23&27	\\
		(N-4) to (N-4)&$\Delta$& & & &  &  & 9& 8& 1& 2	\\	
		(N-4) to (N-4)&$f$		& & & &  &  &24&27&24&29	\\
		(N-5) to (N-5)&$\Delta$& & & &  &  &  &  &11&10	\\
		(N-5) to (N-5)&$f$		& & & &  &  &  &  &35&39	\\
		\hline
	\end{tabular}		
\end{table} 

- With $F=1$, the system is equivalent to a 8-process system, which tolerates $f=6$ in delivering 8 messages to 8 processes (Table 3), tolerates additional $+f=5$ in delivering 7 messages to 7 processes, tolerates additional $+f=3$ in delivering 6 messages to 6 processes (Table 4, Pattern 3), and tolerates additional $+f=1$ in delivering 5 messages to 5 processes (Table 4, Pattern 3), so the resulting tolerance is $f=15$.

- With $F=0$, the system is a 9-process system, which tolerates $f=7$ in delivering 9 messages to 9 processes (Table 3), tolerates additional $+f=6$ in delivering 8 messages to 8 processes (Table 4, Pattern 3), additional $+f=4$ in delivering 7 messages to 7 processes, additional $+f=2$ in delivering 6 messages to 6 processes(Table 4, Pattern 3), and additional $+f=8$ in delivering 5 messages to 5 processes (Table 4, Pattern 4), so the resulting tolerance is $f=27$.

Tolerance of an 8-process system to faulty links per actual number of faulty processes $F$ can be computed (with $F=3$, $F=2$, and $F=1$) using data from Table 3 and (with $F=0$) using data from Table 4 in the following manner:

- With $F=3$, an 8-process system is equivalent to a 5-process system, which tolerates $f=3$ in delivering 5 messages to 5 processes (Table 3).

- With $F=2$, the system is equivalent to a 6-process system, which tolerates f=4 in delivering 6 messages to 6 processes (Table 3), and tolerates additional $+f=3$ in delivering 5 message to 5 processes (Table 3), so the resulting tolerance is $f=7$.

- With $F=1$, the system is equivalent to a 7-process system, which tolerates $f=5$ in delivering 7 messages to 7 process (Table 3), tolerates additional $+f=4$ in delivering 6 messages to 6 processes (Table 3), and tolerates $+f=2$ in delivering 5 messages to 5 processes (Table 3), so the resulting tolerance is $f=11$.

- With $F=0$, it is an 8-process system, which tolerates $f=6$ in delivering 8 messages to 8 processes (Table 1), tolerates additional $+f=5$ in delivering 7 messages to 7 processes (Table 3), tolerates additional $+f=3$ in delivering 6 messages to 6 processes (Table 4, Pattern 3), and tolerates additional $+f=1$ in delivering 5 messages to 5 processes (Table 4, Pattern 3), so the resulting tolerance is $f=15$.

\textit{Note}: It might appear strange that an 8-process system with no faulty processes tolerates $f=15$, while a 7-process system with no faulty processes tolerates $f=17$. 
This outcome illustrates why an additional process can make achieving an agreement harder, yet with no contribution to the tolerance to link faults or asynchrony. Consensus of an 8-process system requires delivery of 5 messages across 5 correct processes, while consensus of a 7-process system requires delivery of 4 messages across 4 correct processes.

\begin{figure}[h]
	\scalebox{0.122}
	{\includegraphics{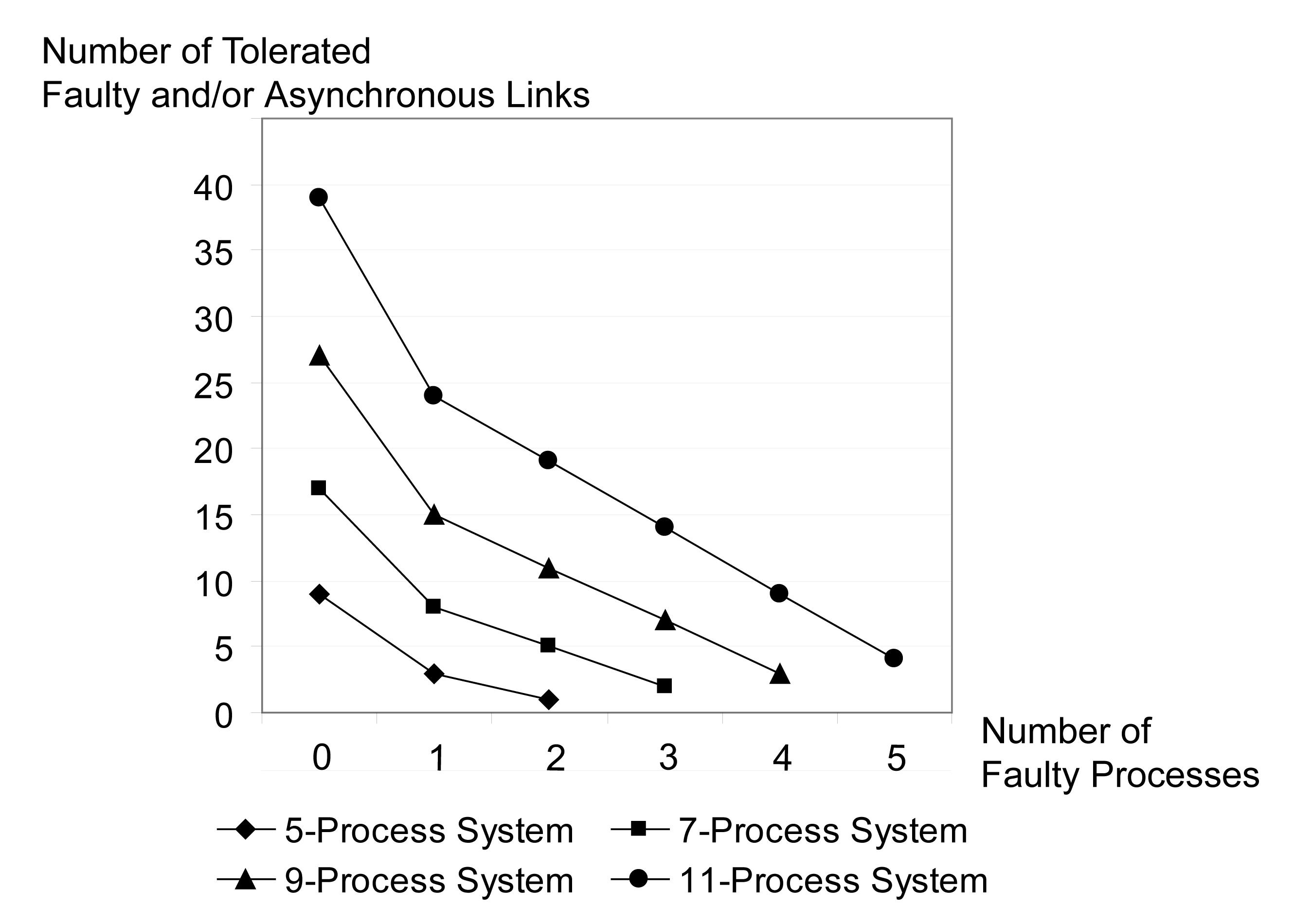}}
	\caption{Tolerance to Async/Faults (Computed + Analytical) }
\end{figure}

\subsection{Tolerance Curves with Extrapolation}

Figure 5 depicts a family of curves, where a curve represents the tolerance of a consensus system to faulty and/or asynchronous links per number of tolerated faulty processes. Tolerance of a 5-process system, a 7-process system, and a 9-process system with 4 and with 3 faulty processes is computed entirely with the simulator.

Tolerance of a 9-process system with 2 faulty processes, with 1 faulty process, and with no faulty process is obtained analytically with extrapolation. Tolerance of an 11-process system is obtained entirely analytically with extrapolation.

\subsection{Conditions for Bounded Termination}

The following equations present tolerance of consensus system to the sum of faulty and asynchronous links $f$ as function of system connectivity $c$ (Graph theory meaning of term connectivity \cite{Deo1974}) and the number of faulty processes $F_{t}$ at a point of time $t$:		

- With odd total number of processes and no faulty processes:

\begin{equation}
	\label{eq:8.1}
	\tag{8.1}
	\begin{Large}
		f < c + c/2 + (c/2)^2
	\end{Large}
\end{equation}

- With odd total number of processes and $F_{t}$ faulty processes:

\begin{equation}
	\label{eq:8.2}
	\tag{8.2}
	\begin{Large}
		f_{t} < c/2 + (c/2)^2 - F_{t}*c/2
	\end{Large}
\end{equation}	

- With even total number of processes and no faulty processes:

\begin{equation}
	\label{eq:8.3}
	\tag{8.3}
	\begin{Large}
		f < (c+1)/2 + ((c+1)/2)^2 - (c+1)/2
	\end{Large}
\end{equation}	

- With even total number of processes and $F_{t}$ faulty processes:

\begin{equation}
	\label{eq:8.4}
	\tag{8.4}
	\begin{Large}
		f_{t} < (c+1)/2 + ((c+1)/2)^2 - (F_{t}+1)(c+1)/2
	\end{Large}
\end{equation}

\vspace{\baselineskip}

These equations are the conditions for bounded termination.
	
\end{document}